\documentclass[%
twocolumn,
superscriptaddress,
 amsmath,amssymb,
 aps,prl
]{revtex4}

\usepackage{amsmath}
\usepackage{amssymb}
\usepackage{wasysym}											%
\usepackage{graphicx}
\usepackage{natbib}												%
\usepackage{fancyhdr}											%
\usepackage{mathtools}										%
\usepackage[usenames,dvipsnames]{xcolor}  %
\usepackage[utf8]{inputenc}               %
\usepackage{multirow}                     %
\usepackage{makecell}                     %
\usepackage[overload]{empheq}             %
\usepackage{ulem}

\definecolor{amaranth}{rgb}{0.9, 0.17, 0.31}
\begin{document}

\title{Observation of magnetic helicoidal dichroism with extreme ultraviolet light vortices}

\author{Mauro Fanciulli\footnote{Authors contributed equally.}}
\email{mauro.fanciulli@u-cergy.fr}
\affiliation{Universit\'e Paris-Saclay, CEA, CNRS, LIDYL, 91191 Gif-sur-Yvette, France}
\affiliation{Laboratoire de Physique des Mat\'eriaux et Surfaces, CY Cergy Paris Universit\'e, 95031 Cergy-Pontoise, France}

\author{Matteo Pancaldi$^*$}
\affiliation{Elettra-Sincrotrone Trieste S.C.p.A., 34149 Basovizza, Trieste, Italy}

\author{Emanuele Pedersoli}
\affiliation{Elettra-Sincrotrone Trieste S.C.p.A., 34149 Basovizza, Trieste, Italy}

\author{Mekha Vimal}
\affiliation{Universit\'e Paris-Saclay, CEA, CNRS, LIDYL, 91191 Gif-sur-Yvette, France}

\author{David Bresteau}
\affiliation{Universit\'e Paris-Saclay, CEA, CNRS, LIDYL, 91191 Gif-sur-Yvette, France}

\author{Martin Luttmann}
\affiliation{Universit\'e Paris-Saclay, CEA, CNRS, LIDYL, 91191 Gif-sur-Yvette, France}

\author{Dario De Angelis}
\affiliation{Elettra-Sincrotrone Trieste S.C.p.A., 34149 Basovizza, Trieste, Italy}

\author{Primo\v{z} Rebernik Ribi\v{c}}
\affiliation{Elettra-Sincrotrone Trieste S.C.p.A., 34149 Basovizza, Trieste, Italy}

\author{Benedikt Rösner}
\affiliation{Paul Scherrer Institut, 5232 Villigen-PSI, Switzerland}

\author{Christian David}
\affiliation{Paul Scherrer Institut, 5232 Villigen-PSI, Switzerland}

\author{Carlo Spezzani}
\affiliation{Elettra-Sincrotrone Trieste S.C.p.A., 34149 Basovizza, Trieste, Italy}

\author{Michele Manfredda}
\affiliation{Elettra-Sincrotrone Trieste S.C.p.A., 34149 Basovizza, Trieste, Italy}

\author{Ricardo Sousa}
\affiliation{Universit\'e Grenoble Alpes, CNRS, CEA, Grenoble INP, IRIG-SPINTEC, 38000 Grenoble, France}

\author{Ioan-Lucian Prejbeanu}
\affiliation{Universit\'e Grenoble Alpes, CNRS, CEA, Grenoble INP, IRIG-SPINTEC, 38000 Grenoble, France}

\author{Laurent Vila}
\affiliation{Universit\'e Grenoble Alpes, CNRS, CEA, Grenoble INP, IRIG-SPINTEC, 38000 Grenoble, France}

\author{Bernard Dieny}
\affiliation{Universit\'e Grenoble Alpes, CNRS, CEA, Grenoble INP, IRIG-SPINTEC, 38000 Grenoble, France}

\author{Giovanni De Ninno}
\affiliation{Elettra-Sincrotrone Trieste S.C.p.A., 34149 Basovizza, Trieste, Italy}
\affiliation{Laboratory of Quantum Optics, University of Nova Gorica, 5001 Nova Gorica, Slovenia}

\author{Flavio Capotondi}
\affiliation{Elettra-Sincrotrone Trieste S.C.p.A., 34149 Basovizza, Trieste, Italy}

\author{Maurizio Sacchi}
\affiliation{Sorbonne Universit\'e, CNRS, Institut des NanoSciences de Paris, INSP, 75005 Paris, France}
\affiliation{Synchrotron SOLEIL, L’Orme des Merisiers, Saint-Aubin, B. P. 48, 91192 Gif-sur-Yvette, France}

\author{Thierry Ruchon}
\affiliation{Universit\'e Paris-Saclay, CEA, CNRS, LIDYL, 91191 Gif-sur-Yvette, France}

\begin{abstract}
We report on the experimental evidence of magnetic helicoidal dichroism, observed in the interaction of an extreme ultraviolet vortex beam carrying orbital angular momentum with a magnetic vortex. Numerical simulations based on classical electromagnetic theory show that this dichroism is based on the interference of light modes with different orbital angular momenta, which are populated after the interaction between the light phase chirality and the magnetic topology. This observation gives insight into the interplay between orbital angular momentum and magnetism, and sets the framework for the development of new analytical tools to investigate ultrafast magnetization dynamics.
\end{abstract}
\maketitle

Beyond plane waves, light beams may feature helical wavefronts, with the Poynting vector precessing with time around the beam's propagation axis \cite{Allen1992}. 
The number of intertwined helices spiralling clockwise or anticlockwise defines the topological charge $\ell\in\mathbb{Z}$, which is associated to the orbital angular momentum (OAM) of the light vortices. This is independent from their polarization state, which instead is associated to a spin angular momentum (SAM) \cite{BliokhPR2015}. 
OAM light beams are nowadays harnessed for an ever increasing scope of applications covering different fields from microscopy \cite{Klar:1999,Wei:2015, Tamburini:2006} and biology \cite{Paterson:2001, Grier:2003}, to telecommunications \cite{Fickler:2016, Gong:2019} and quantum technologies \cite{Yao2011, Shen2019}. 
Vortex beams also play a role in spectroscopy, where the coupling between the OAM and the internal degrees of freedom of atoms, atomic ions or molecules has been exploited to transfer OAM to these species \cite{Schmiegelow2016, Afanasev:2018, Forbes:2018, DeNinno:2020}, and to enhance enantiomeric sensitivity \cite{Brullot2016, Ni:2021}. 
Also, a rich variety of examples arises for the investigation and manipulation of topologically complex objects such as chiral magnetic structures \cite{Sirenko:2019} and skyrmions \cite{Fujita:2017, Fujita:2017b, Yang:2018}. 
In the same way as tuning the wavelength is used to achieve chemical contrast, or tuning the polarization to achieve magnetic contrast, controlling the OAM state of a vortex beam has the potential to provide topological contrast in systems possessing a well-defined handedness. 
This general statement can eventually find applications in many different XUV and X-ray based techniques, like elastic or inelastic scattering and photoelectron emission. The study of magnetic structures is a particularly appealing case, for their practical importance and for the possible control of their topology. 

Over the last decade, the development of highly coherent sources and tailored optical schemes has opened new possibilities for generating structured light vortices in the extreme ultraviolet (XUV) \cite{Gariepy2014,GeneauxNcom2016,Kong2017,Gauthier2017,Ribic2017,Dorney:2019,Rego2019,Pisanty:2019} and x-ray \cite{Hemsing2013,Vila-Comamala:2014,Sasaki:2015,Lee:2019,Loetgering:2020,Woods:2021} regimes, paving the way to their spectroscopic applications. 
In this context, magnetic helicoidal dichroism (MHD) has been recently predicted \cite{Fanciulli:2021A}, in analogy to the SAM dependent magnetic circular dichroism (MCD). 
Upon interaction (reflection or transmission) of a pure Laguerre-Gaussian mode of topological charge $\ell$ with a magnetic surface, MHD consists in an intensity redistribution into all modes $\ell+n$ in the outgoing beam, where $n$ represents all the azimuthal decomposition coefficients of the magnetic structure topological symmetry. \cite{Fanciulli:2021A}. 
Differently from MCD,  MHD is sensitive to the overall topology of the spin texture, it vanishes for homogeneous structures, and is not self-similar if one inverts either the helicity of the beam or the magnetization direction. 

Among a great variety of magnetic structures in two \cite{Zhang2020} or three dimensions \cite{FernandezPacheco2017}, magnetic vortices (MV) are particularly promising for technological applications \cite{Shinjo2000,Bader2006}. 
They can form in mesoscopic dots that are much larger than their thickness, leading to an in-plane curling magnetization \cite{Natali:2002}. 
They have been shown to be particularly robust against perturbations \cite{Braun2012} and present a rich sub-nanosecond dynamics \cite{Guslienko2006}. Because of their symmetry, MVs are also a particularly simple test case for MHD, since they only present $n=\pm1$ \cite{Fanciulli:2021A}.

In this letter, we report on the experimental observation of MHD by measuring the resonant scattering of XUV radiation carrying OAM from a permalloy (Py, Fe$_{20}$Ni$_{80}$) dot in MV micromagnetic configuration. We compare the experimental results to theoretical predictions \cite{Fanciulli:2021A} and we interpret them in terms of the interference between the different $\ell+n$ modes of the reflected light. Our study illustrates the potential of MHD as a new optical tool for the investigation of magnetic structures.\\

\begin{figure}[!htp]
\centering
\includegraphics[width=0.5\textwidth]{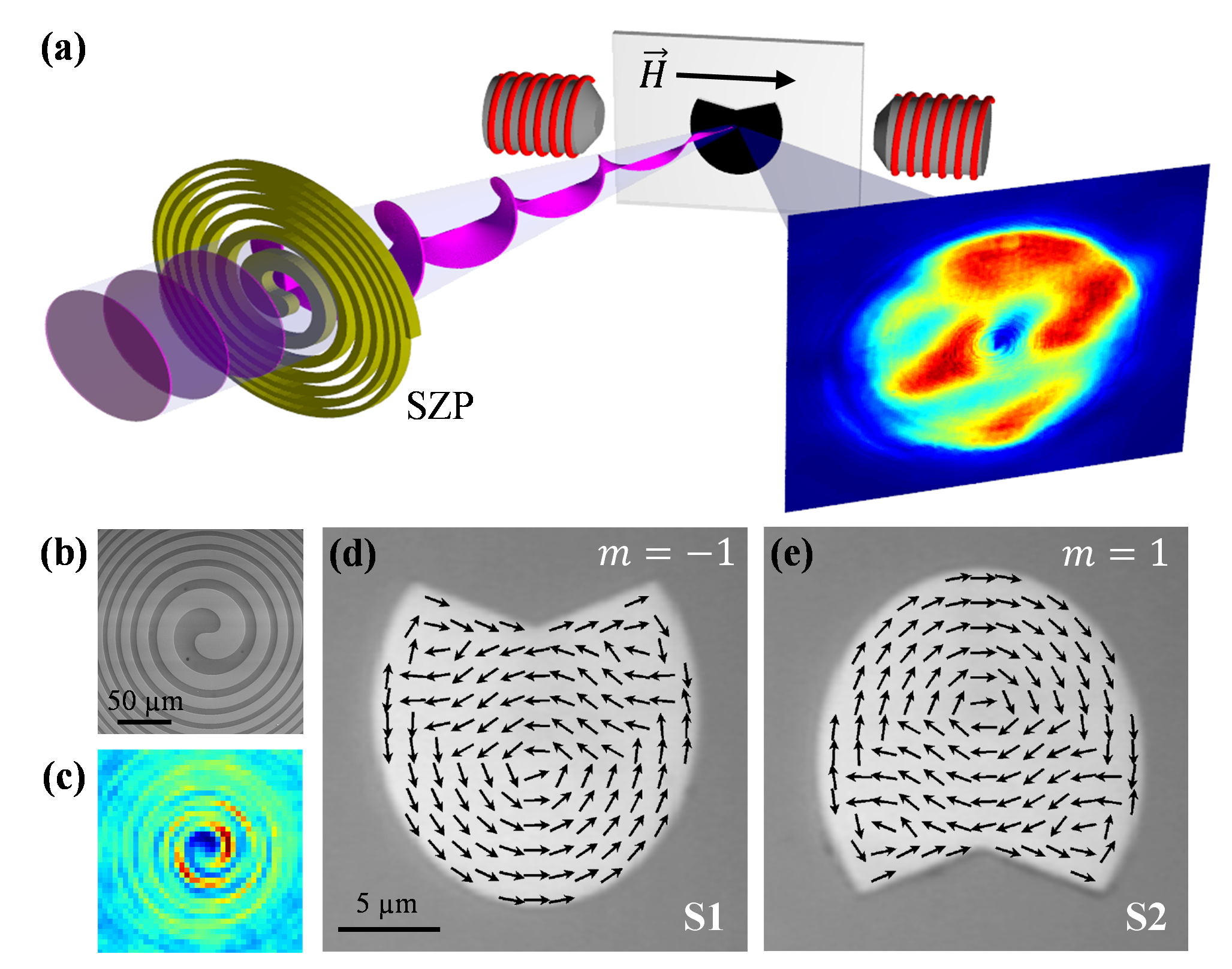}
\caption{(a) Schematics of the experimental setup at the DiProI beamline, showing the incoming FEL beam with planar wavefront, a SZP that imparts OAM to the FEL beam, the sample placed between the poles of the electromagnet and the image of the scattered beam; 
(b) scanning electron microscope image of one SZP ($\ell=+1$); 
(c) far-field image of the interference pattern of the corresponding OAM beam and the undiffracted, OAM-free beam, featuring a spiraling intensity pattern;
(d)-(e) magnetic dots $S1$ and $S2$ with corresponding remanent magnetization after applying a pulse of saturating external field $\vec{H}$ as in (a). An opposite field reverses the remanent MV chirality in both samples. \label{Fig1}}
\end{figure}
The experiment was performed at the DiProI beamline \cite{Capotondi:2013} of the FERMI free-electron laser (FEL) \cite{Allaria:2012} using the setup sketched in Fig.~\ref{Fig1}(a). 
The spatially coherent close-to-Gaussian FEL beam is focused on the sample by one of three available silicon zone plates \cite{SOM:} 
mounted on a movable stage. One is a Fresnel zone plate producing a focused beam with $\ell=0$, the other two are spiraling zone plates (SZP) that impart OAM to the beam with either $\ell=-1$ or $\ell=+1$. The properties of the zone plates are detailed in \cite{Ribic2017}. An electron microscope image of the SZP for $\ell=+1$ is shown in Fig.~\ref{Fig1}(b), while in (c) the direct image of the beam with OAM collected by a CCD camera shows the expected spiral shaped far-field interference pattern with the undiffracted beam \cite{Fuerhapter2005, Ribic2017}.

The linearly $P$-polarized XUV beam has an energy of $52.8$~eV, matching the Fe $3p\!\rightarrow\!3d$ core resonance \cite{Valencia2006}, and impinges on the center of the sample at an angle of $48^\circ$ from the normal, which is close to the Brewster extinction condition, in order to enhance the magneto-optical effects. Using a knife edge scan, we measured a spot size of about $4~\mu$m (full width half maximum) at the sample plane, in agreement with previous characterizations \cite{Rosner:2017}. The reflected beam is collected by a CCD camera. 

The samples [$S1$ and $S2$, Fig~\ref{Fig1}(d),(e)] are two identical and $\pi$-rotated ellipsoidal Py dots with a triangular indent, prepared on the same Si substrate. They are $80$~nm thick, their short diameter is $15~\mu$m and they are protected by a $\approx3$~nm Al layer (oxidized in air) \cite{SOM:}. 
Their exact shape was optimized by micromagnetic calculations \cite{oommf} in order to satisfy two criteria: i) feature at remanence a single stable MV with a diameter larger than the XUV beam spot size; ii) enable the switching of the remanent vortex chirality $m=\pm1$ by in situ application of a moderate external magnetic field pulse. The calculated remanent magnetization after a $+20$~mT in-plane magnetic pulse [arrow in Fig.~\ref{Fig1}(a)] shows MVs of opposite chirality for $S1$ and $S2$ [Fig.~\ref{Fig1}(d)-(e)], providing a simple way of cross-checking our experimental results. Reversing the magnetic pulse direction switches the chirality of both MVs. 
Further experimental details are given in \cite{SOM:}.\\
\begin{figure*}[!htp]
\centering
\includegraphics[width=1\textwidth]{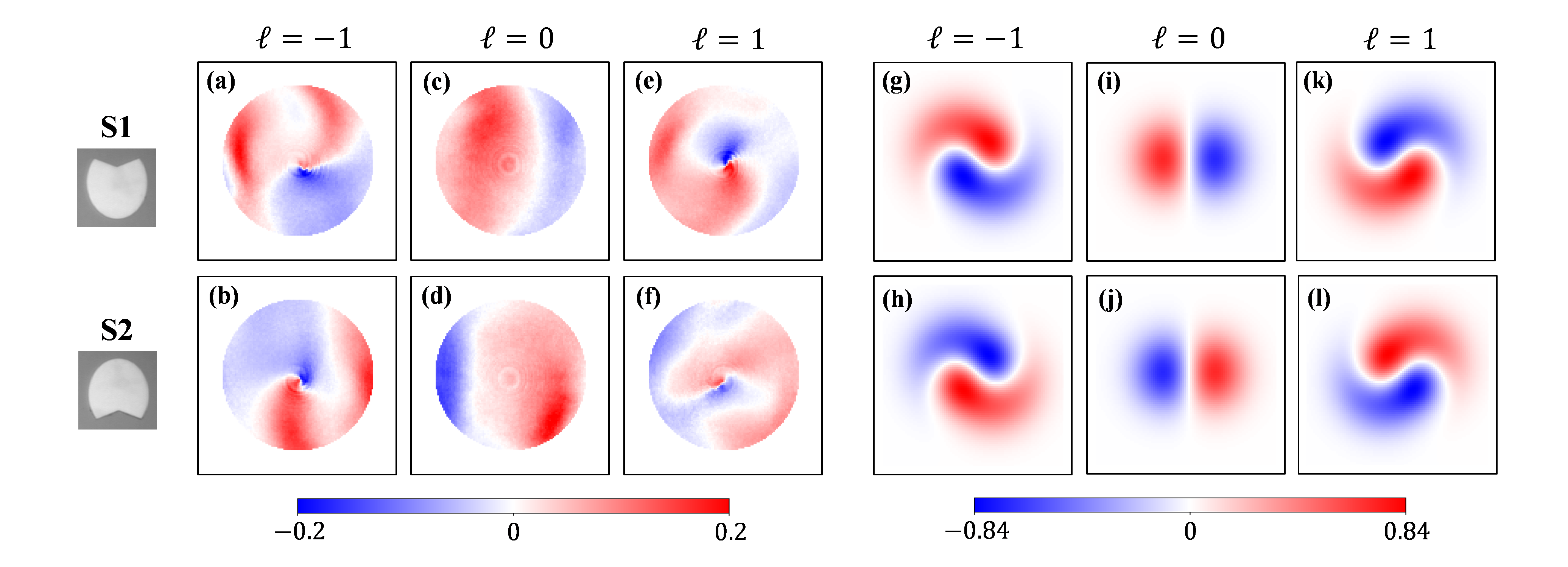}
\caption{(a)-(f) Experimental dichroism (corresponding to MHD$m$) for $\ell=-1,0,+1$ for MVs in dot $S1$ (top) and $S2$ (bottom). (g)-(l): Numerical simulations for the same experimental parameters. \label{Fig2}}
\end{figure*}
For each sample, we evaluate the dichroic signal by switching the sign of the external magnetic field pulse before measuring at remanence, so that no mechanical or optical adjustment of the setup is required, guaranteeing optimal stability in the measurement conditions. 
Also, in this way non-magnetic contributions to the scattered
intensity are largely suppressed from the difference signal. Fig.~\ref{Fig2}(a)-(f) shows the experimental dichroism on the intensity profile of the reflected beam for three incoming $\ell$ values and for the two samples. Details of the data analysis are given in \cite{SOM:}. 
For sample $S1$, we observe a left-right asymmetry for $\ell=0$ [Fig.~\ref{Fig2}(c)], and opposite spiral asymmetries for opposite topological charges of the OAM beam $\ell=\pm1$ [Fig.~\ref{Fig2}(a),(e)], showing the differential dependence on the topological charge of the OAM beam. 
The measured MHD signal is of the order of $20\%$. 
The result is reproduced in $S2$ [Fig.~\ref{Fig2}(b),(d),(f)], where the color pattern is reversed since the two samples always have opposite $m$. 
This demonstrates the magnetic nature of the observed dichroic signal rather than possible asymmetries induced by the experimental setup.  
It is worth noting that the symmetry relations observed in Fig.~\ref{Fig2}(a)-(f) are cancelled when the MV topology is perturbed by applying an external magnetic field, as reported in \cite{SOM:}.

\begin{figure}[!htp]
\centering
\includegraphics[width=0.5\textwidth]{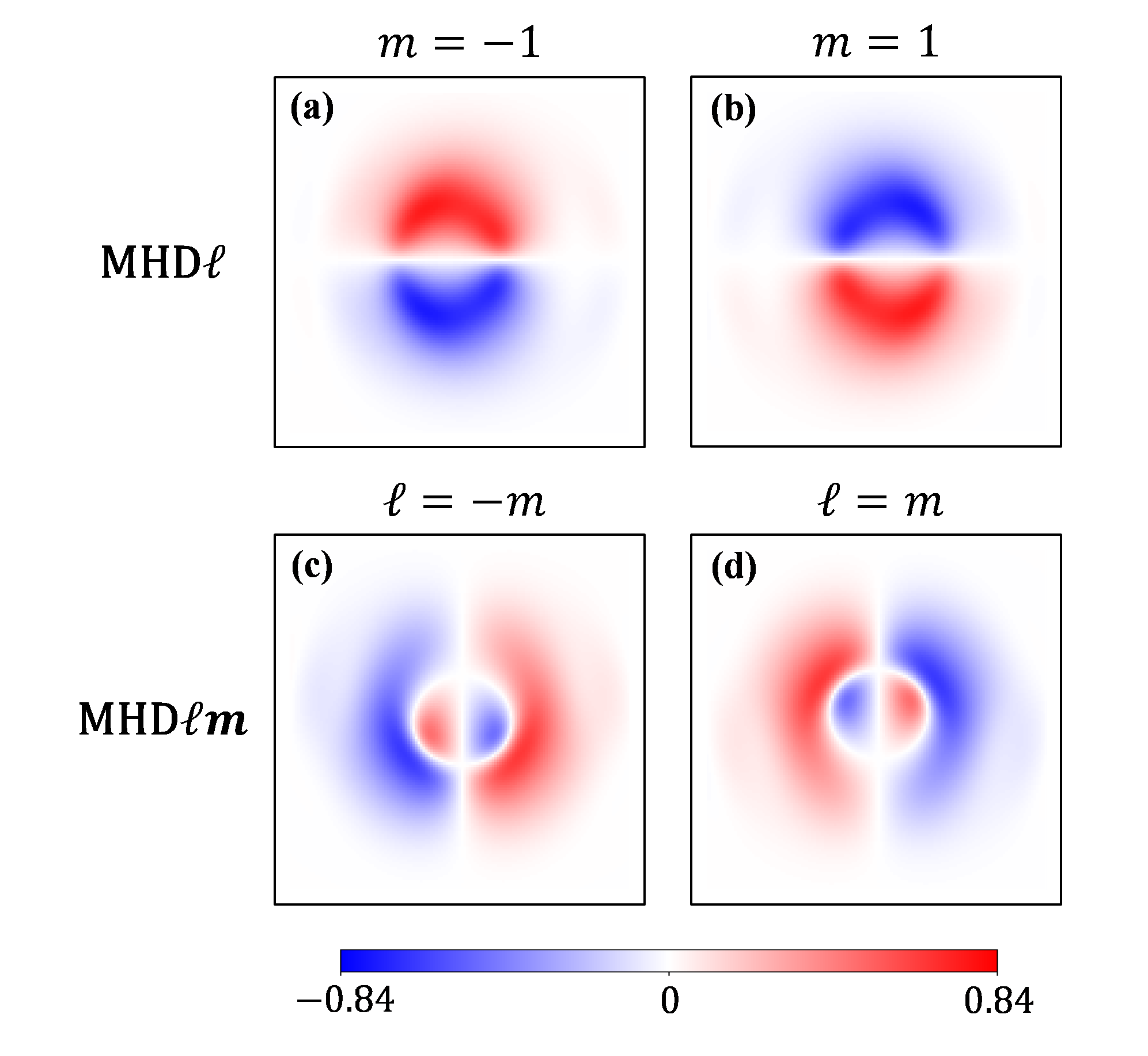}
\caption{Numerical simulations with same experimental parameters of (a)-(b) MHD$\ell$ for different $m$, and (c)-(d) MHD$\ell m$ for different combinations of $\ell$ and $m$. \label{Fig3}}
\end{figure}
In order to interpret the experimental results, we implemented a numerical model for MHD based on Ref.~\cite{Fanciulli:2021A}, with a perfect gaussian beam as input, ideal optics and perfect centering of the beam on the MV. The results of the simulations for the same scattering geometry as in the experiment are shown in Fig.~\ref{Fig2}(g)-(l). They can be directly compared with panels (a)-(f), showing a good agreement for all configurations. Quantitatively, the experimental dichroism is lower than expected, mainly because of the non-magnetic signal coming from the oxidized Al-capping layer not taken into account in the simulations \cite{SOM:}. 
Here it is important to stress that the dependence of the MHD signal on the topological charge is not linked to the particular chosen reflection geometry close to the Brewster angle. From further simulations \cite{SOM:} we predict that even close to normal incidence the asymmetry for $\ell=0$ (left-right) is different from $\ell=\pm1$ (top-bottom), while the spiral asymmetry of Fig.~\ref{Fig2} is due to a combination with geometric effects \cite{Fanciulli:2021A}.\\

In general, in order to evaluate the MHD, three relevant combinations of differences between signals obtained with given values of topological charge of the beam and chirality of the MV can be considered \cite{SOM:}.
Defining $I_{\ell,m}$ the far field intensity of the reflected beam, we classify these three kinds of dichroisms as \cite{Fanciulli:2021A}:
\begin{subequations}
\begin{align}[left = \empheqlbrace\,]
\text{MHD}\ell &=\left(I_{\ell,m}-I_{-\ell,m}\right)/\left(I_{\ell,m}+I_{-\ell,m}\right)\\
\text{MHD}m &=\left(I_{\ell,m}-I_{\ell,-m}\right)/\left(I_{\ell,m}+I_{\ell,-m}\right) \label{mhdm}\\
\text{MHD}\ell m &=\left(I_{\ell,m}-I_{-\ell,-m}\right)/\left(I_{\ell,m}+I_{-\ell,-m}\right)\,.
\label{eq:MHD}
\end{align}
\end{subequations}
The dichroism presented in Fig.~\ref{Fig2} corresponds to MHD$m$, since switching the magnetic field direction corresponds to switching the chirality of the MV for a fixed value of $\ell$.
In Fig.~\ref{Fig3} the other two kinds of MHD from simulations are shown. The dichroic intensity map of MHD$m$ [Fig.~\ref{Fig2}(g),(k)] differs from MHD$\ell$ [Fig.~\ref{Fig3}(a)-(b)], which ensures a nonzero MHD$\ell m$ [Fig.~\ref{Fig3}(c)-(d)]. This is a crucial difference with respect to MCD, where switching the sample magnetization or the light polarization is equivalent. 

The two cases of MHD$\ell$ for a given $m$ [Fig.~\ref{Fig3}(a)-(b)] can be exchanged by time reversal (i.e. switching the MV chirality) \cite{SOM:}, and the same is true for MHD$\ell m$ for a given combination of $\ell$ and $m$ [Fig.~\ref{Fig3}(c)-(d)]. On the contrary, the two MHD$m$ cases of opposite $\ell$ [Fig.~\ref{Fig2}(g),(k)] are exchanged through parity inversion (i.e. switching the helicity $\ell$ of the OAM) \cite{SOM:}, and correspond to a truly chiral configuration \cite{Barron2009,Barron:2013}. Indeed, the MHD$m$ images are chiral patterns that cannot be superimposed by rotation, while the MHD$\ell$ and MHD$\ell m$ ones can be exchanged by a $\pi$ rotation. 
From the experimental point of view, while changing $m$ is done with an external magnetic field pulse, changing $\ell$ requires a mechanical displacement of the SZPs, preventing an easily reproducible beam alignment on the sample and making 
the experimental observation of MHD$\ell$ and MHD$\ell m$ less reliable in our setup.\\ 


In order to complement the theoretical analysis proposed in \cite{Fanciulli:2021A}, it is interesting to look into the physical mechanism at the origin of MHD$m$. 
Since the curling magnetization of the MV determines the geometry of the magneto-optical constants, the reflectivity matrix will depend on the azimuth $\phi$. This can be used to intuitively retrieve the simple selection rule $\Delta\ell=\pm1$ for reflection by a MV \cite{Fanciulli:2021A}. In fact the magnetization terms of the MV, and hence the reflectivity coefficients, will vary as $\cos\left(\phi\right)$ up to a constant phase term, while the azimuthal dependence of the incoming electric field due to the OAM will read $\cos\left( -\frac{2\pi}{\lambda}z-\ell\phi\right)$, with $\lambda$ and $z$ being the light wavelength and the propagation direction, respectively. Without considering geometric effects \cite{Fanciulli:2021A}, the magnetic contribution to the outgoing electric field is given by the product of those two terms and will thus show $(\ell\pm 1)\phi$ components, and only these, while the non-magnetic one is only given by a reflection of the $\ell$ mode. The propagation of the $\ell$ and $\ell\pm1$ modes to the far field will result in interferences in the intensity profile, which is what causes the asymmetries of MHD. Note that this is valid for every incoming $\ell$, including $\ell=0$. 

\begin{figure}[!htp]
\centering
\includegraphics[width=0.4\textwidth]{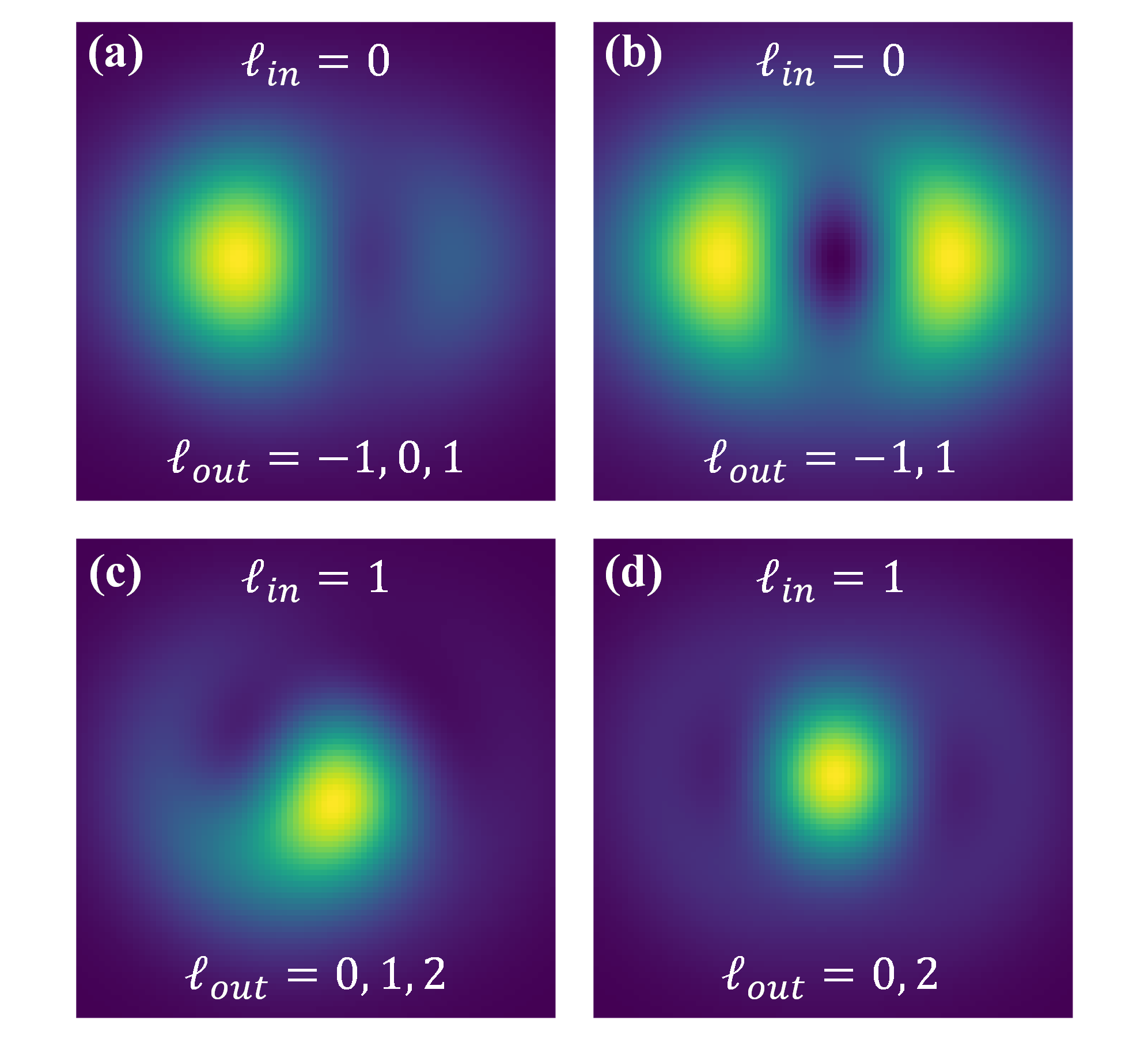}
\caption{(a) Far field intensity of a beam with $\ell_{in}=0$ reflected by a MV with $m=+1$; (b) same as (a) when the outgoing mode $\ell_{out}=\ell_{in}$ is suppressed. (c) and (d) present the same results as (a) and (b) when $\ell_{in}=1$. \label{Fig4}}
\end{figure}

To be more specific, let us consider a $P$-polarized beam with $\ell_{in}=0$, that is a standard Gaussian plane wave without OAM. 
Fig.~\ref{Fig4}(a) shows the far-field intensity after reflection from a $m=+1$ MV, featuring a left-right imbalance that will reverse for the case $m=-1$ \cite{SOM:}, leading to MHD$m$ [Fig.~\ref{Fig2}(i)]. 
If we suppress the propagation of the outgoing mode $\ell_{out}=\ell_{in}=0$ in the simulation and leave only $\ell_{out}=\pm1$ we obtain the intensity in Fig.~\ref{Fig4}(b). This is the typical shape of a Hermite-Gaussian beam of index $1$. We can draw two important conclusions. 
One is that even a standard Gaussian beam with $\ell=0$ reflected by a chiral magnetic structure such as a MV contains equal weights of modes with opposite non-zero OAM. This is a largely overlooked observation that can be used as a fresh way to look at magneto-optical scattering phenomena. The second conclusion is that the dichroic signal in MHD originates from the interference of the main incoming mode $\ell_{out}=\ell_{in}$ with the newly generated ones. This can also be seen for example for the case of $\ell_{in}=1$. In Fig.~\ref{Fig4}(d) where $\ell_{out}=\ell_{in}=1$ is suppressed from the reflected beam one finds signatures of a second order Hermite-Gaussian beam, while the full beam of Fig.~\ref{Fig4}(c) is the result of the interference with the central mode. In this case a spiraling shape is obtained. Its mirrored image on the horizontal axis is obtained when one considers $\ell_{in}=-1$ instead, mirrored on the vertical axis when $\ell_{in}=-1$ and $m=-1$, and mirrored on both axis when $\ell_{in}=+1$ and $m=-1$ \cite{SOM:}. The fact that these four cases are not equivalent explains why MHD$m$ and MHD$\ell$ are different, and thus the existence of MHD$\ell m$.\\

In conclusion, we presented the experimental evidence of magnetic helicoidal dichroism in the resonant reflection of an XUV beam that carries OAM by a magnetic vortex, and found a good agreement with theoretical predictions. 
In particular, we identify the specific dependence of MHD on the sign of the optical ($\ell$) and magnetic ($m$) vortices. 
Other configurations of OAM and light polarization could be explored \cite{SOM:}, providing altogether a new way to look at the magneto-optical scattering process in terms of OAM, as discussed also for the case of a $\ell=0$ incoming beam. 

It is straightforward to extend this approach to other complex magnetic structures. For example, in analogy with the recent observation that infrared vortex beams with opposite $\ell$ are sensitive to dipolar chiral nanohelix \cite{Wozniak:2019, Ni:2021b}, one can envision to use MHD for XUV and soft X-ray beams with OAM in order to detect the helical direction of complex 3D spiral spin structure \cite{Birch:2020}, or to study chiral domain walls and skyrmions in magnetic films with strong Dzyaloshinskii–Moriya interaction \cite{Legrand:2018}. Conversely, different structures can be engineered in order to tailor or analyze the OAM content of a light beam \cite{Lee:2019, Woods:2021}. A limit of our interpretation is that it is based on classical electromagnetism. A microscopic theory has still to be developed and questions about a possible local exchange of OAM between light and matter remain open. 
Finally, the feasibility of this experiment with a FEL source naturally opens up the study of MHD in the time domain, as recently performed in nanoplasmonics \cite{Spektor:2017, Dai:2020}. MHD could then become a powerful experimental tool for the study of the ultrafast dynamics of magnetic materials and of their chiral properties \cite{Kerber:2020, Buttner:2021}, but also as a way to control and manipulate them, in a similar fashion as in the inverse Faraday effect with SAM \cite{Kimel:2005}.\\

This work was supported by: 
the Agence Nationale pour la Recherche (under Contracts No. ANR11-EQPX0005-ATTOLAB and No. ANR14-CE320010-Xstase); 
the Swiss National Science Foundation project No. P2ELP2\_181877; 
the Plateforme Technologique Amont de Grenoble, with the financial support of the CNRS Renatech network; 
the EU-H2020 Research
and Innovation Programme, No. 654360 NFFA-Europe.\\

\bibliographystyle{unsrtmauro}

\end{document}